%% file: MasterArxiv.tex
\documentclass[aps,showpacs,superscriptaddress,pra,notitlepage,twocolumn]{revtex4-1}
\usepackage{amsmath}
\usepackage{amsfonts}
\usepackage{amssymb}
\usepackage[caption=false]{subfig}
\usepackage{appendix}
\usepackage{graphicx}
\usepackage{braket}
\usepackage{xcolor}
\usepackage{hyperref}
\usepackage{floatrow}

\newcommand{\fig}[1]{\mbox{Fig.\ \ref{#1}}}
\newcommand{\eqn}[1]{\mbox{Eq.\ \ref{#1}}}

\begin{document}

\input{HeptagonCode}

\appendix
\section{Supplemental Material}
\input{SuppMat}

\end{document}

%% file: HeptagonCode.tex
  \title{Calderbank-Steane-Shor Holographic Quantum Error Correcting Codes}
  \author{Robert J. Harris}
  \email{rjh2608@gmail.com}
\affiliation{ARC Centre for Engineered Quantum Systems, School of Mathematics and Physics, The University of Queensland, St Lucia, QLD, 4072, Australia}
\author{Nathan A. McMahon}
\affiliation{Center for Engineered Quantum Systems, Dept. of Physics and Astronomy, Macquarie University, 2109 New South Wales, Australia}
\affiliation{ARC Centre for Engineered Quantum Systems, School of Mathematics and Physics, The University of Queensland, St Lucia, QLD, 4072, Australia}
\author{Gavin K. Brennen}
\affiliation{Center for Engineered Quantum Systems, Dept. of Physics and Astronomy, Macquarie University, 2109 New South Wales, Australia}
\author{Thomas M. Stace}
\affiliation{ARC Centre for Engineered Quantum Systems, School of Mathematics and Physics, The University of Queensland, St Lucia, QLD, 4072, Australia}
  \date{\today}
  
\begin{abstract}
We expand the class of holographic quantum error correcting codes by developing the notion of block perfect tensors, a wider class that includes previously defined  perfect tensors. The relaxation of this constraint opens up a range of other holographic codes. We demonstrate this by introducing the self-dual CSS heptagon holographic code, based on the 7-qubit Steane code. Finally we show promising thresholds for the erasure channel by applying a straightforward, optimal erasure decoder to the heptagon code and benchmark it against existing holographic codes.
\end{abstract} 

\pacs{}

  \maketitle
  
The correspondence between anti-de Sitter (AdS) space and conformal field theories (CFT) \cite{Maldacena1999} is an example of the holographic principle between a bulk $d+1$-dimensional AdS and a boundary $d$-dimensional CFT \cite{Maldacena2003}. 
 AdS space is a maximally symmetric solution to the vacuum Einstein equations, in particular it is a solution with negatively curved spacetime. Boundary CFTs are quantum field theories invariant under conformal transformations. This is currently the most precise realisation of the holographic principle and has spurred much work in this field \cite{Harlow2016}. 
It has been conjectured that any CFT can interpreted as a theory of quantum gravity which is an asmptotically AdS space \cite{Harlow2016}, with an appropriate choice of metric on both sides
. 

A feature of the correspondence is the Ryu-Takayanagi (RT) formula relates the von Neumann entropy of a $d$-dimensional CFT boundary region to the minimal surface area of the $d+1$-dimensional AdS bulk, that subtends the boundary region \cite{Ryu2006,Hubeny2007}. 


This correspondence suggests that the boundary degrees of freedom possess substantial redundancy, making it a candidate for robustly encoding quantum information. 
Holographic codes were first proposed as a way to connect quantum information with the bulk/boundary correspondence \cite{Latorre2015, Yoshida2013}. Here we focus on the construction introduced in Ref.\ \cite{Pastawski2015}, which combines three desirable features: they are stabiliser codes and thus exactly solvable, they are quantum error correction codes (QECC), and their encoding is described by a tensor network which is a uniform tiling of hyperbolic space. \citeauthor{Pastawski2015} demonstrated that a family of holographic codes based on the five qubit QECC satisfies the RT formula (although since the two-point correlators are not scale-invariant the stabiliser codes do not correspond to a CFT on the boundary). A key requirement of \citeauthor{Pastawski2015} is that the network be comprised exclusively of \emph{perfect tensors}, described below, which strongly constrains the  encoding circuit.

 

The erasure thresholds shown for these codes are comparable to the performance of certain tree networks \cite{Varnava2006} and the surface code \cite{Stace2009, Stace2010, Barrett2010}. A favourable comparison between the surface code and the pentagon holographic code suggest this is a promising avenue for practical codes. However there are a number of scenarios where it is beneficial to use Calderbank-Steane-Shor (CSS) codes, for example in building large scale cluster states for measurement based computation \cite{Raussendorf2006,Briegel2009,Rudolph2017}, or for building foliated codes \cite{Bolt2016} to use as long-range quantum repeaters.

 
 

In this letter, we show that perfect tensors are not required for constructing holographic codes and describe a CSS heptagon code based on the seven qubit Steane code. 
We also implement an exact erasure decoder for holographic codes, and demonstrate its performance on the heptagon code and the original pentagon code. The optimal decoder outperforms the greedy algorithm from \cite{Pastawski2015}, and gives a threshold of $\sim1/3$ for the heptagon code.

\begin{figure}[!t]
\includegraphics[width=\columnwidth]{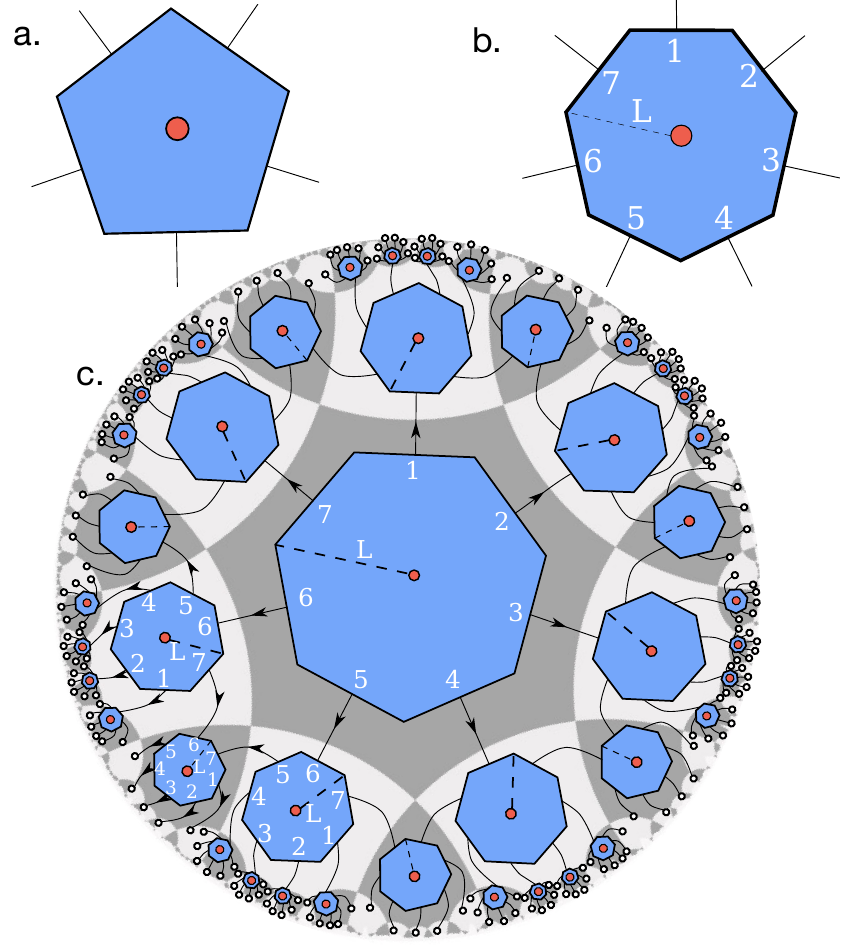}
\caption{(a) Graphical representation of the seed tensor for the 5-qubit pentagon code. The central red dots represent a logical input qubit.  (b) Graphical representation of the seed tensor for the Steane 7-qubit heptagon code.  The cyclic ordering of indices ensures a block perfect Steane code presentation as in \eqn{eq:SteaneTensor}. (c) Tessellation of the heptagonal seed in a larger tensor network representation of radius $R=3$ heptagon code. The small circles around the boundary  represent physical qubits. The arrows show the conventional direction from the centre to the boundary, which we adopt for constructive purposes. Numbered indices indicate a consistent ordering required to produce an isometry from the bulk logical qubits to boundary physical qubits. 
}
\label{fig:tensors}\label{fig:pentagontensor}\label{fig:heptagontensor}\label{fig:heptagontiling}
\end{figure}

\citet{Pastawski2015} construct holographic quantum error correcting codes based  on the \mbox{$[[n,k,d]]=[[5,1,3]]$} 5-qubit code \cite{Laflamme1996}. The 5-qubit code generates a rank $n+1=6$ \emph{seed tensor}, $T$, represented graphically as a pentagon in \fig{fig:pentagontensor}a. 
%
%
%
%
%
The central vertex represents a logical input qubit, and the $n$ planar legs represent output qubits. The tensor specifies an isometry from logical input operators to output operators. 

A larger tensor network is built from the seed tensor by tessellating it in the $\{4,n\}$ Schl\"afli geometry (i.e.\ with four polygons meeting at each vertex), forming a surface with negative curvature illustrated in \fig{fig:heptagontiling}c.  Neighbouring indices on adjacent tensors are contracted, which we represent graphically by connecting the corresponding planar legs \cite{Bridgeman2017}. 
The tessellation terminates at a certain radius $R$, which is given by the minimal number of edges from the boundary to the central \emph{bulk} logical qubit.  Input  vertices on each seed tensor in the bulk represent  logical bulk qubits; the uncontracted legs at the boundary terminate at  physical  qubits, denoted by hollow circles.

The seed tensors in the pentagon holographic code are \emph{perfect tensors}, which we briefly review. For a rank $2m$  tensor, $T$, we may partition its indices into an ordered set $A$ and its ordered complement ${\bar A}$ such that \mbox{$\lvert A \rvert \leq \lvert {\bar A} \rvert=2m-\lvert A \rvert$}. We interpret $T$ as a linear map from the logical Hilbert space on the input indices in set $A$ to the image Hilbert space on the output indices in set ${\bar A}$, i.e.\
$
T^{{\bar A} \leftarrow A} : \mathcal{H}_A \mapsto \mathcal{H}_{{\bar A}}.
$

\citet{Pastawski2015} define $T$ to be a perfect tensor if it is an isometry for \emph{all} bipartitions of the indices  $\{A | {\bar A}\}$ with $\lvert A \rvert \leq \lvert {\bar A} \rvert$, i.e.\ $T$ satisfies
\begin{equation}
\big(T^{{\bar A} \leftarrow A}\big)^\dagger T^{{\bar A} \leftarrow A} = \mathbb{I}_{\mathcal{H}_A}.
\label{eq:perfectdef}
\end{equation}
This is a very restrictive constraint on $T$.

For later discussion, we note that we can express a given a bipartition of indices $\{A|{\bar A}\}$ as a permutation $\Pi$ with respect to some reference index ordering $\underline J=\{j_1,j_2,...,j_{2m}\}$, i.e.\ \mbox{$\{A|{\bar A}\}= \Pi[\underline J]$}. A perfect tensor is therefore an isometry for \emph{all} permutations $\Pi$. This formulation will help when we define the less restricted class of \emph{block perfect tensors}.

Operators acting on the physical qubits on the boundary of the space are defined via \emph{operator pushing} \cite{Pastawski2015} from logical bulk qubits, through the tensor network, to the physical boundary qubits. For constructive purposes we assign a direction to each leg in the network, shown by arrows in \fig{fig:heptagontiling}c, indicating that an `output' index from one tensor contracts with an `input' index to an adjacent tensor. Each tensor in the network translates operators acting on input indices to operators acting on output indices, according to 
\begin{equation}
O^A=T^{A\leftarrow {\bar A}}O^{{\bar A}} \left(T^{A\leftarrow {\bar A}}\right)^\dagger.
\label{eq:tensorpushing}
\end{equation}
For example, an identity operator on the logical input leg of a stabiliser code translates to any of the stabiliser group on the output legs.

We note that the bulk logical qubits near the boundary map, via operator pushing, to boundary operators that are localised on a small wedge of the physical boundary qubits, while logical operators for qubits deep within the bulk are highly delocalised over the boundary \cite{SuppMat}.

The perfect tensor property guarantees that tensor legs can be arbitrarily partitioned into inputs and outputs. While this was inspired by diffeomorphism invariance in the underlying AdS space \cite{Pastawski2015}, it is very restrictive.    
They show that the rank-6 seed tensor for the 5-qubit code is indeed a perfect tensor. 
The resulting holographic code has finite rate $r=k/n=1/\sqrt{5}$ in the asymptotic limit. 


We now show 
that the perfect tensor constraint can be relaxed to a less restrictive class that still generates a holographic code.  This is motivated by the observation that in the hyperbolic tessellation in \cite{Pastawski2015}, input legs to a given seed tensor can be grouped into a contiguous block.  This new class includes the seven qubit Steane code \cite{Steane1996} tensor, which we use in a CSS holographic code based on a  tiling of heptagons.

We define \emph{block perfect tensors} to be those that are isometries for all \emph{cyclic} permutations $\Pi=\sigma^p$ of $\underline J$, i.e.\ those for which  $\{A|{\bar A}\}=\sigma^p[\underline J]$, where $\sigma^p:j_i\mapsto j_{i+p}$ is a cyclic shift. This coincides with the description of perfect tangles developed independently \cite{Berger2018}. We believe this constraint on the seed tensor more closely corresponds to a discretisation of diffeomorphism invariance, though we do not comment on this further here. 





\emph{Steane Tensor}: We exemplify this relaxation of perfection by showing that the rank-8 Steane tensor, $T^{j_1,...,j_L,j_7}_{\mathrm{Steane}}$, which is generated from the  Steane code, is block perfect, but not perfect.  This tensor is defined through the unique simultaneous $+1$ eigenstate, $\ket{T_{\mathrm{Steane}}}$, of the 8 stabilisers
\begin{equation}
\begin{array}{rlllllllll}
\textrm{index label:} & 1 & 2 & 3 & 4 & 5 & 6 & L & 7 \\ 
 \hline
 & X & X & \mathbb{I} & \mathbb{I} & \mathbb{I} & X & \mathbb{I} & X & \equiv S_1 \\ 
& \mathbb{I} & X & X & X & \mathbb{I} & \mathbb{I} & \mathbb{I} & X& \equiv S_2 \\ 
 & \mathbb{I} & \mathbb{I} & \mathbb{I} & X & X & X & \mathbb{I} & X& \equiv S_3 \\ 
 & Z & Z & \mathbb{I} & \mathbb{I} & \mathbb{I} & Z & \mathbb{I} & Z & \equiv S_4\\ 
 & \mathbb{I} & Z & Z & Z & \mathbb{I} & \mathbb{I} & \mathbb{I} & Z & \equiv S_5\\ 
& \mathbb{I} & \mathbb{I} & \mathbb{I} & Z & Z & Z & \mathbb{I} & Z & \equiv S_6\\
 & X & X & X & X & X & X & X & X & \equiv S_{\bar X}\\ 
& Z & Z & Z & Z & Z & Z & Z & Z & \equiv S_{\bar Z}
\end{array},
\label{eq:SteaneTensor}
\end{equation}
via the Choi-Jamio\l kowski isomorphism \cite{Jiang2013}.
That is, $T^{j_1,...,j_L,j_7}_{\mathrm{Steane}}=\langle j_1,...,j_L,j_7\ket{T_{\mathrm{Steane}}}$ where  $\ket{T_{\mathrm{Steane}}}$ satisfies  $S_\alpha\ket{T_{\mathrm{Steane}}}=\ket{T_{\mathrm{Steane}}}$ for all $\alpha$. The index labels in \eqn{eq:SteaneTensor} are consistent with the ordering shown in figure \fig{fig:heptagontensor}b. With respect to this index label ordering, we have exhaustively checked that $T_{\mathrm{Steane}}$ is block perfect.


It is straightforward to see that $T_{\mathrm{Steane}}$ is \emph{not} a perfect tensor, by considering the non-contiguous partition of indices $A=\{3,4,5,L\}$ and ${\bar A}=\{1,2,6,7\}$. If $T_{\mathrm{Steane}}$ were perfect, then $T^{A\leftarrow {\bar A}}$ would be unitary. Then according to \eqn{eq:tensorpushing}, \mbox{$T^{A\leftarrow {\bar A}}X^{\otimes A} \big(T^{A\leftarrow {\bar A}}\big)^\dagger\neq\mathbb{I}^{\otimes {\bar A}}$}. However according to $S_1$, and the Choi-Jamio\l kowski isomorphism
\mbox{$T^{A\leftarrow {\bar A}}X^{\otimes A} \big(T^{A\leftarrow {\bar A}}\big)^\dagger = \mathbb{I}^{\otimes {\bar A}}$}, implying that for this partition, $T^{A\leftarrow {\bar A}}$ is not a unitary map.  Hence $T_{\mathrm{Steane}}$ is not perfect.

\emph{Heptagon Holographic Code}: As with the pentagon code, the heptagon code is built on a 2D tiling with negative curvature. This is a tessellation of heptagons, with four heptagons meeting at each vertex (the $\{4,7\}$ Schl\"afli geometry), as shown in \fig{fig:heptagontiling}c.

The hyperbolic tiling of the heptagon code requires a consistent assignment of  index contractions between adjacent tensors. \fig{fig:heptagontiling}c shows one such assignment for a subset of the tiles. This ensures that every seed tensor indeed acts as an isometry from inputs to outputs, so that the entire network is an isometry from bulk inputs to boundary outputs. 

%





Because the seed code is a self-dual CSS code, it is clear that pushing $X$-like tensors will lead to $X$-like holographic stabilisers, and similarly for the $Z$-like stabilisers. This means the heptagon holographic code is a self-dual CSS code. Similarly to the pentagon code, the heptagon code is a finite rate code, with asymptotic rate  $r=1/\sqrt{21}$.

Block-perfect tensors in a hyperbolic tiling generate a holographic code according to the definition in \citeauthor{Pastawski2015}, based on the existence of a greedy algorithm.  
The greedy algorithm constructs a recoverability region, $\mathcal{R}$, of bulk logical data by recursively adding bulk tensors, $T$ to $\mathcal{R}$ according to the local update rules: (1) boundary qubits are `recoverable'�� if they are not erased; and (2) given some set, ${\bar A}$, of tensor indices for $T$ that are recoverable��, then if there is an isometry from $A$ to ${\bar A}$  (with $\lvert A \rvert \leq \lvert {\bar A} \rvert$), then we add $T$ to $\mathcal{R}$. The fixed point of these rules defines $\mathcal{R}$. For holographic codes built from perfect tensors, the tensor indices in ${\bar A}$ can be arbitrary, whereas for  block perfect tensors they must be in contiguous order. Starting from a contiguous region $B$ on the boundary, the region $\mathcal{R}$ produced by this algorithm has an inner boundary that approximates, to within a small constant, the discrete bulk geodesic $\gamma_B$ connecting the end points of B. As shown in \cite{Pastawski2015} this implies an RT formula of $S_B \propto \lvert\gamma_B\rvert$.


\emph{Erasure decoders}: Having defined the heptagon holographic code we are interested in the resilience of the code to errors. \citet{Delfosse2016} propose using the robustness of a code to erasure errors as a proxy for performance of the code under more general error channels. As such, we now describe an erasure decoder for this code, with which to quantify the code performance.

Loss errors are heralded, so that we know where they have occurred. This enables us to use the error pattern as part of the error decoding algorithm, making an exact decoder computationally feasible. 

\emph{Recovery Algorithm}:
The algorithm we detail here is optimal for any stabiliser code, including the holographic codes. 
Computationally, it relies on matrix row reduction, which for an $a\times b$ dimensional matrix has run time $\sim O(a^2 b)$ (there are more sophisticated algorithms with lower complexity \cite{Bunch1974}).
In the optimal decoder $a$ is the number of erasure errors and $b=n-k$ is the number of stabilisers.

For simplicity we describe the algorithm for CSS codes, however it is straightforwardly adapted to any stabiliser code. The stabilisers for an $[[n,k,d]]$ CSS code are specified by a set of binary support vectors $\underline{s}_j$ such that the $X$-like stabilisers are given by $S_j=\hat{X}^{\otimes \underline{s}_j}$. Likewise a logical support vector $\underline{\ell}$ defines an $X$-like logical operator $\bar X=X^{\otimes \underline{\ell}}$ \cite{Gottesman1997}. $Z$-like stabilisers and logical operators are defined similarly.

Logical operators are equivalent up to multiplication by stabilisers, so that $\bar X'=X^{\otimes \underline{\ell}'}\sim\bar X=X^{\otimes \underline{\ell}}$  iff
\begin{equation}
\underline{\ell}^\prime=\underline{\ell} + {\sum}_j \lambda_j \underline{s}_j \mod{2},
\end{equation}
for some $\lambda_j \in \mathbb{Z}_2$.

Suppose a subset of physical qubits are erased.  This error is  defined by a binary support vector $\underline{\varepsilon}$, in which an entry 1 in position $i$ indicates that the $i^{\textrm{th}}$ qubit is lost. Providing we can construct a logical operator $X^{\otimes \underline{\ell}'}$ which has no support on the lost qubits, i.e.\ $ \underline{\ell}^\prime \cdot \underline{\varepsilon} = 0$, then  the corresponding logical information is recoverable 
(note that the dot product here is not modular).

It is clear that to satisfy this condition it is necessary and sufficient to find $\underline{\ell}^\prime$ which has zeros at positions where $\underline{\varepsilon}$ is 1, i.e.\ ${\ell}^\prime_i=0$ if $\varepsilon_i=1$. We define a \emph{filtered} support vector, $\underline{a}^{(\underline{\varepsilon})}$, which is the restriction of the support vector $\underline{a}$ to the positions at which $\underline{\varepsilon}=1$. 
Then $\underline{\ell}^\prime \cdot \underline{\varepsilon} = 0$ iff we can find ${\lambda_j}$ s.t.:
\begin{equation}
\underline{\ell}^{(\underline{\varepsilon})}+{\sum}_j \lambda_j \underline{s}_j^{(\underline{\varepsilon})}=0 \mod{2}.
\label{eq:recoverycriteria}
\end{equation}
The existence (and solution where required) of satisfying ${\lambda_j}$'s can be determined efficiently with row reduction of the matrix of filtered stabiliser support vectors augmented with the filtered logical support vector.

\emph{Monte-Carlo Simulations}:
To evaluate the performance of this decoder, and the performance of the heptagon code under erasure, we simulate the recovery of the central logical qubit after loss using Monte-Carlo simulations. We generate i.i.d patterns of physical qubit erasure for a fixed number of errors $a=\mathrm{wt}(\underline{\varepsilon})$, and then use the algorithm detailed above to determine whether each pattern is recoverable. We iterate over all $a\in\{0,1,..., n\}$ to estimate the recovery probability, $P_{\mathrm{rec}}(a,n)$, and use the binomial formula
\begin{equation}
p_{\textrm{rec}}(p, n)={\sum}_a {n\choose a} p^a (1-p)^{n-a} P_{\mathrm{rec}}(a,n),
\label{eq:binomial}
\end{equation}
to calculate the recovery rate for different loss rates $p$.

\begin{figure}[!t]
\includegraphics[width=0.95\textwidth]{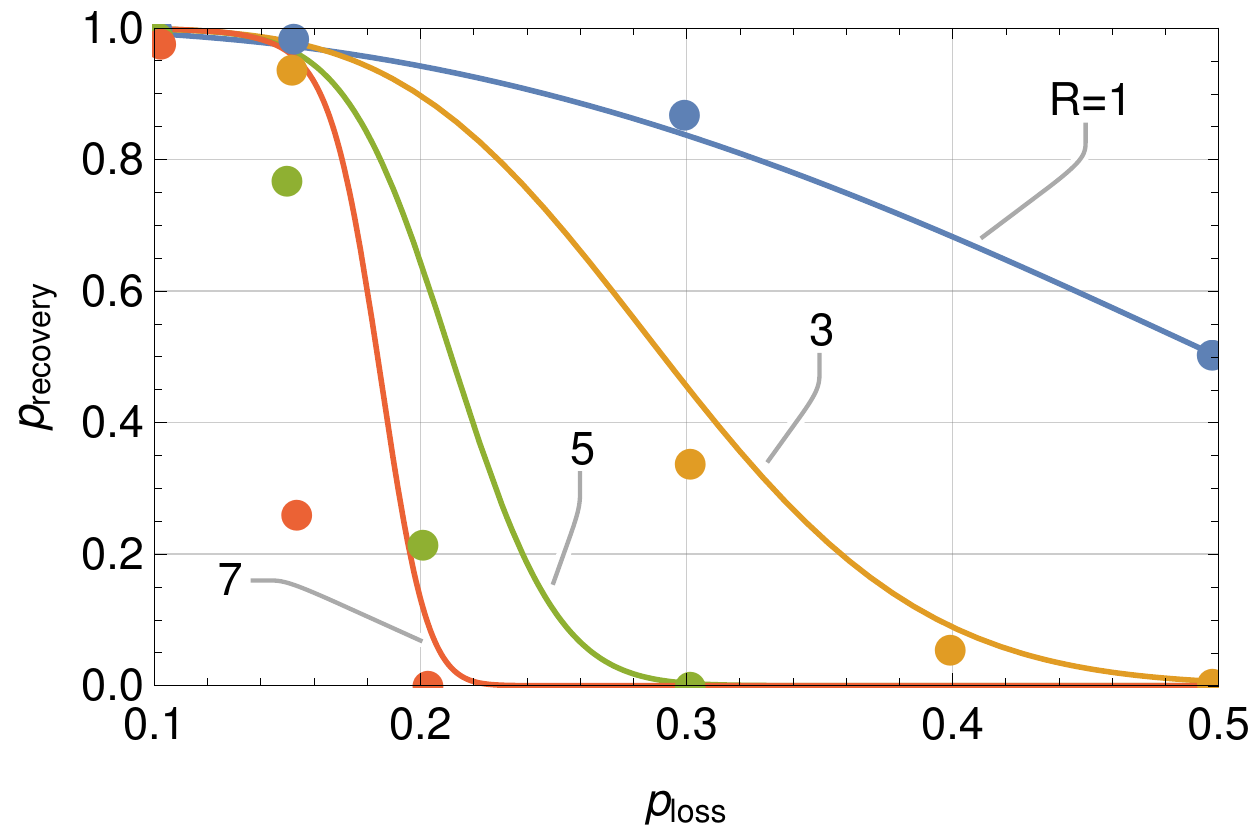}
\caption{Recovery probability for the pentagon code, based on recovery of the central logical qubit using a greedy algorithm  (points) \cite{Pastawski2015}, and the optimal row reduction algorithm calculated according to as in \eqn{eq:binomial} (solid lines) for codes of radius \mbox{$R=1,3,5,7$}. Note that the $R=1$ code is the five qubit code. As $R$ grows, the optimal decoder performs increasingly better then the greedy decoder.}
\label{fig:PentagonRecovery}
\end{figure}

We use the pentagon code to benchmark the row reduction algorithm against the greedy algorithm in \cite{Pastawski2015}. This code does not have an erasure threshold as the distance of the central qubit of code does not increase with the radius. The results are shown in \fig{fig:PentagonRecovery} where lines are the results of numerical simulations with the optimal decoder and \eqn{eq:binomial}, and points are from the heuristic greedy algorithm in \cite{Pastawski2015}. As the radius of the network increases, a growing discrepancy between the optimal row-reduction and the heuristic greedy algorithm is evident, albeit with no threshold appearing.  

\begin{figure}[htp]
\includegraphics[width=0.95\textwidth]{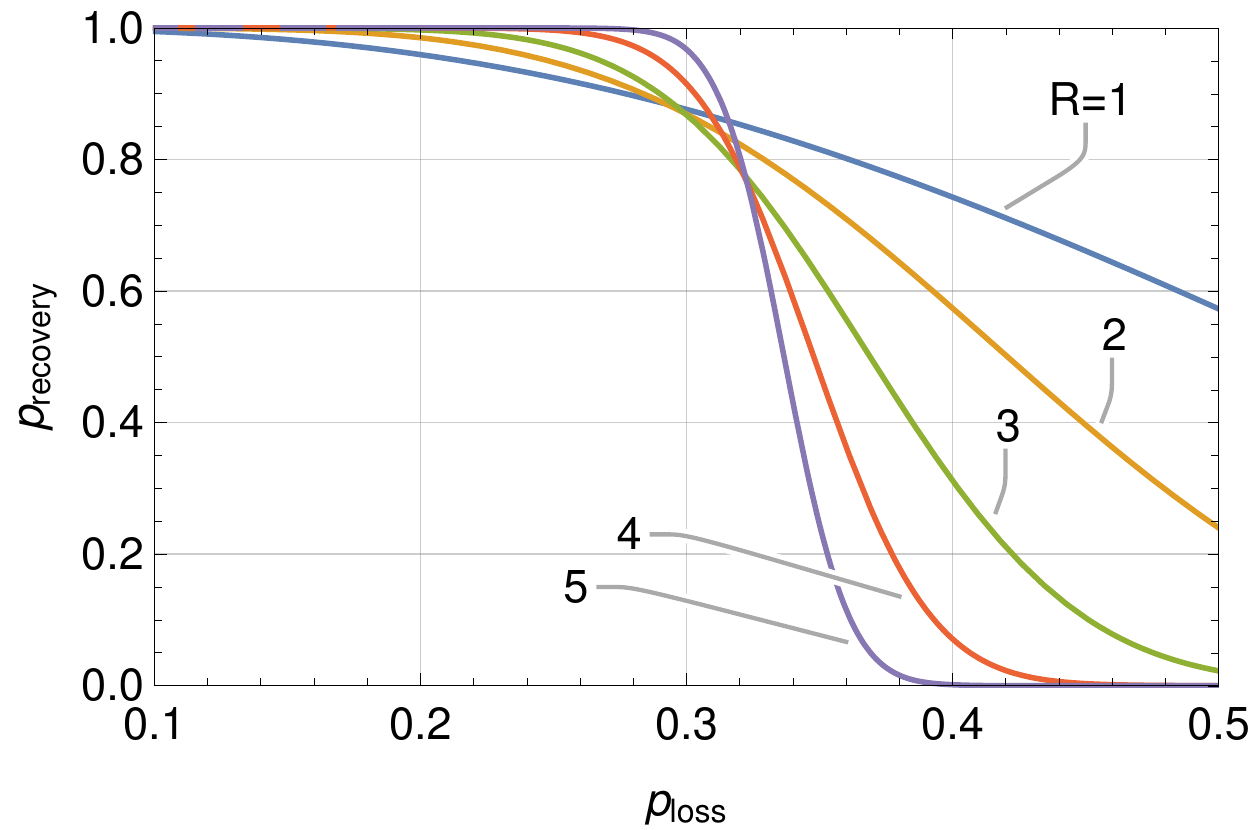}
\caption{Recovery probability for the heptagon code, \mbox{$p_{\mathrm{rec}}(p_{\mathrm{loss}})$} as in \eqn{eq:binomial}, for recoverability of the central logical qubit for codes of  radius \mbox{$R=1,2,3,4,5$}. Note that the $R=1$ code is the Steane code.  There is a threshold at $p_{\mathrm{loss}}<p^*_{\mathrm{hept}}$. }
\label{fig:HeptagonRecovery}
\end{figure}

We now examine the performance of the heptagon holographic code against erasure, as measured by the recovery probability for the central logical qubit. The performance curves are shown in \fig{fig:HeptagonRecovery} up to a radius $R=5$ code. In contrast to the pentagon code, we do find a threshold, $p^*_{\mathrm{hept}}\approx 1/3$ for erasure in the heptagon code. That is, for an erasure probability $p_{\mathrm{loss}}<p^*_{\mathrm{hept}}$, the code performance improves with increasing radius.

We compare the performance of the heptagon code to the mixed pentagon/hexagon code of \citeauthor{Pastawski2015}, where the asymptotic rate is reduced by a factor of around $1/2$, to find a threshold of around \mbox{$p^*_{\mathrm{pent/hex}}\approx 1/3$}. We calculate the asymptotic rate of their pentagon/hexagon code to be \mbox{$r_{\mathrm{pent/hex}}=(13 \sqrt{6}-12)/90\approx 1/\sqrt{20.5}$}, which is very similar to the rate for the heptagon code $r_\textrm{hept}=1/\sqrt{21}$ proposed here. 

The high erasure threshold and code rate suggest that the heptagon code might be an attractive candidate for a practical error correcting code in networks that have weak geometric constraints, such as optical architectures \cite{Rudolph2017}.  Its performance against other logical error channels is the subject of ongoing research.  Further, since it is of CSS form, the heptagon code can be constructed from measurements on a suitably prepared cluster state \cite{Bolt2016}.  However, because some stabilisers have relatively large weight, the cluster states resulting from that construction would usually be high-valence, which tends to amplify imperfections during cluster creation.  We note here that the heptagon holographic code can be implemented with a low-valence cluster state; details will be reported in a forthcoming publication.  

To conclude, we have developed the notion of block perfect tensors, a less restricted class than the perfect tensors introduced in earlier work.  This makes  a range of other codes available for tessellation in a  holographic tensor network, including the self-dual CSS Steane code, with which we have exemplified the general construction.  Finally, we have applied a straightforward, optimal erasure decoder based on matrix row reduction on filtered support vectors to characterise the performance of holographic codes, yielding promising thresholds.

\emph{Acknowledgements}:
We appreciate discussions with Miguel J B Ferreira and Tobias Osborne.
This work was supported by the Australian Research Council Centre of Excellence for Engineered Quantum Systems (Grant No. CE 110001013).
  \bibliography{/home/uqrhar15/Dropbox/Documents/Bibtex/HolographicCodes}

%% file: SuppMat.tex
  \title{SUPPLEMENTAL MATERIAL: \\ Calderbank-Steane-Shor Holographic Quantum Error Correcting Codes}
  \author{Robert J. Harris}
  \email{rjh2608@gmail.com}
\affiliation{ARC Centre for Engineered Quantum Systems, School of Mathematics and Physics, The University of Queensland, St Lucia, QLD, 4072, Australia}
\author{Nathan A. McMahon}
\affiliation{Center for Engineered Quantum Systems, Dept. of Physics and Astronomy, Macquarie University, 2109 New South Wales, Australia}
\affiliation{ARC Centre for Engineered Quantum Systems, School of Mathematics and Physics, The University of Queensland, St Lucia, QLD, 4072, Australia}
\author{Gavin K. Brennen}
\affiliation{Center for Engineered Quantum Systems, Dept. of Physics and Astronomy, Macquarie University, 2109 New South Wales, Australia}
\author{Thomas M. Stace}
\affiliation{ARC Centre for Engineered Quantum Systems, School of Mathematics and Physics, The University of Queensland, St Lucia, QLD, 4072, Australia}
  
  \maketitle
  
  Given a particular code associated to each of the seed tensors, there is a 
  unique choice of stabilisers where each stabiliser is centred on a particular seed tensor. Each stabiliser can be considered to stabilise a particular logical input.
  
  In \fig{fig:heptagontiling}c there is a direction associated to each physical leg on the seed tensors. For the central seed tensor all physical legs are outputs, for all other seed tensors there is either one or two input physical legs.  

Starting with seed tensors at the boundary we produce a subset of stabilisers for the holographic code associated to these boundary seed tensors. This is done by selecting the stabilisers which have identity on all input legs. For example, using the Steane code as the seed tensor, the stabilisers associated to each boundary seed tensor with identity on index six and logical are:
\begin{equation}
\begin{array}{rccccccccl}
&&&&&&\multicolumn{2}{c}{\mathrm{Input}}&\\[-2.5mm]
&&&&&&\multicolumn{2}{c}{$\downbracefill$}&\\[-1mm]
\textrm{index label:} & 1 & 2 & 3 & 4 & 5 & 6 & L & 7 \\ 
 \hline
 & \mathbb{I} & X & X & X & \mathbb{I}& \mathbb{I}& \mathbb{I} & X  & \equiv S_2 \\
 & X & X & \mathbb{I} & X & X & \mathbb{I} & \mathbb{I}& \mathbb{I}  & \equiv S_1 S_3 \\
 & \mathbb{I} & Z & Z & Z & \mathbb{I}& \mathbb{I}& \mathbb{I} & Z  & \equiv S_5 \\
 & Z & Z & \mathbb{I} & Z & Z & \mathbb{I}& \mathbb{I}& \mathbb{I}   & \equiv S_4 S_6 \\
 \end{array}.
\label{eq:StabilisersSingleIncoming}
\end{equation}
A stabiliser for the holographic code, associated with the boundary seed tensors, is the operators above on the output physical qubits from the particular seed tensor along with identity on all other physical qubits. As alternate choices from \eqn{eq:StabilisersSingleIncoming} come from the Steane code generating set, they create independent stabilisers for the holographic code. So all options from \eqn{eq:StabilisersSingleIncoming}, with identities on all other physical qubits can be added to the generating set for the heptagon code.

Since different boundary seed tensors do not share physical qubits, the procedure above will generate independent holographic stabilisers for each boundary seed tensor. For the same reason they are all guaranteed to commute. This procedure generates a subset of the stabiliser generators for the holographic code.

Adding the additional restriction of identities on two planar legs (indices six and seven), we have the stablisers: 
%
\begin{equation}
\begin{array}{rccccccccl}
&&&&&&\multicolumn{3}{c}{\mathrm{Input}}&\\[-2.5mm]
&&&&&&\multicolumn{3}{c}{$\downbracefill$}&\\[-1mm]
\textrm{index label:} & 1 & 2 & 3 & 4 & 5 & 6 & L & 7 \\ 
 \hline
 & X & X & \mathbb{I} & X & X & \mathbb{I}& \mathbb{I}& \mathbb{I} & \equiv S_1 S_3 \\
 & Z & Z & \mathbb{I} & Z & Z & \mathbb{I}& \mathbb{I}& \mathbb{I} & \equiv S_4 S_6 \\
 \end{array}.
\label{eq:StabilisersDoubleIncoming}
\end{equation}
As for the stabilisers from \eqn{eq:StabilisersSingleIncoming}, these can produce further stabilisers for the generating set, which again are independent and commuting for the same reasons as above.

%

For non-boundary seed tensors, the approach above constitutes the initial stage of generating the stabilisers. From this, the operators are pushed to the boundary to form the holographic stabiliser.

Operator pushing is essentially taking a seed tensor and inputting either an $X$ or $Z$ operator for each incoming leg. We can use then use stabiliser $S_{1} S_{2}$ or $S_{4}S_{5}$ to pull these onto all other legs:
\begin{equation}
\begin{array}{rccccccccc}
&&&&&&\multicolumn{2}{c}{\mathrm{Input}}&\\[-2.5mm]
&&&&&&\multicolumn{2}{c}{$\downbracefill$}&\\[-1mm]
\textrm{index label:} & 1 & 2 & 3 & 4 & 5 & 6 & L & 7 \\ 
 \hline
 & X & \mathbb{I} & X & X & \mathbb{I} & X & \mathbb{I} & \mathbb{I} & \equiv S_{1} S_{2}  \\
 & Z & \mathbb{I} & Z & Z & \mathbb{I} & Z & \mathbb{I} & \mathbb{I}  & \equiv S_{4} S_{5}  \\
 \end{array},
\label{eq:PushingSingleIncoming}
\end{equation}
This push is unique upto product of stabilisers (with identity on legs L and 6) of the seed tensor we are pushing this operator through.

In the normal operator pushing language we have the 
seed tensor $T$, which is proportional to an isometry from input physical legs (and logical leg) to outgoing physical legs and can define operator $O^{'}$ from $O$ by:
\begin{equation}
TO \propto TO ( T^{\dagger} T) = ( TO T^{\dagger}) T = O^{'}T
\end{equation}
Note operators are only defined upto multiplication by stabilisers.

When there are multiple incoming legs then we can do the same using the appropriate choice from seed tensors:
\begin{equation}
\begin{array}{rccccccccl}
&&&&&&\multicolumn{3}{c}{\mathrm{Input}}&\\[-2.5mm]
&&&&&&\multicolumn{3}{c}{$\downbracefill$}&\\[-1mm]
\textrm{index label:} & 1 & 2 & 3 & 4 & 5 & 6 & L & 7 \\ 
 \hline
 & X & \mathbb{I} & X & X & \mathbb{I} & X & \mathbb{I} & \mathbb{I} & \equiv S_{1} S_{2} \\
 & \mathbb{I} & X & X & X & \mathbb{I} & \mathbb{I} & \mathbb{I} & X & \equiv S_{2} \\
  & X & X & \mathbb{I} & \mathbb{I} & \mathbb{I} & X & \mathbb{I} & X & \equiv S_{1} \\
  & Z & \mathbb{I} & Z & Z & \mathbb{I} & Z & \mathbb{I} & \mathbb{I} & \equiv S_{4} S_{5} \\
 & \mathbb{I} & Z & Z & Z & \mathbb{I} & \mathbb{I} & \mathbb{I} & Z & \equiv S_{5} \\
   & Z & Z & \mathbb{I} & \mathbb{I} & \mathbb{I} & Z & \mathbb{I} & Z & \equiv S_{4}  \\
 \end{array}.
\label{eq:PushingDoubleIncoming}
\end{equation}

Note all identity inputs, push through to all identity outputs.

Repeated applications of operator pushing is sufficient to push the operators from the non-boundary seed tensor to the boundary physical qubits. This generates an further subset of non-boundary stabilisers that completes the stabiliser generating set for the holographic code.

This procedure generates six stabilisers for the central tensor, four stabilisers for each tensor with one input physical leg, and two stabilisers for two input physical legs. This can be shown to sum to the number of required stabilisers. As they are all commuting and independent this is a stabiliser generating set.